# Quantum-Circuit-Based Visual Fractal Image Generation in Qiskit and Analytics

Hillol Biswas
*Department of Electrical and Computer Engineering*
*Democritus University of Thrace*
Xanthi, Greece

*Abstract*— As nature is ascribed as quantum, the fractals also pose some intriguing appearance which is found in many micro and macro observable entities or phenomena. Fractals show self-similarity across sizes; structures that resemble the entire are revealed when zoomed in. In Quantum systems, the probability density or wavefunction may exhibit recurring interference patterns at various energy or length scales. Fractals are produced by basic iterative rules (such as Mandelbrot or Julia sets), and they provide limitless complexity. Despite its simplicity, the Schrödinger equation in quantum mechanics produces incredibly intricate patterns of interference and entanglement, particularly in chaotic quantum systems. Quantum computing, the root where lies to the using the principles of quantum-mechanical phenomenon, when applied in fractal image generation, what outcomes are expected? The paper outlines the generation of a Julia set dataset using an approach coupled with building quantum circuit, highlighting the concepts of superposition, randomness, and entanglement as foundational elements to manipulate the generated dataset patterns. As Quantum computing is finding many application areas, the possibility of using quantum circuits for fractal Julia image generation posits a unique direction of future research where it can be applied to quantum generative arts across various ecosystems with a customised approach, such as producing an exciting landscape based on a quantum art theme.

I. INTRODUCTION

The rise of a new creative form in connection with the advancement of science and computer technology already has a history, as the quantum representation of the world has fundamentally altered perspectives for over a century. In conjunction with the disseminating extensions in many domains as cryptography, medicine, finance, communication, quantum computing is evolving in arts and creativity as well. The topic of quantum algorithmic pictures is strongly impacted by randomness and other probability distribution aspects present in quantum computing. Randomness evolved become an artistic undertaking as well as a scientific tool [1]. While Richard Feynman's postulation of a possible quantum computer [2] for physics simulation transpired four decades back, Schrodinger presented a fresh approach to depicting the world as it is, based on quantum mechanics, on the quantum representation of the world at tiny scales [3]. Rather than using the standard addition, subtraction, multiplication, and division operators, the very basic experiments used logic gates and qubits to encode pictures and three-dimensional objects [4]. Utilizing the concepts of quantum physics as a source of inspiration, quantum art [4] blends art and science. This questionnaire's dynamic context and creative shape, which draws inspiration from the uncertainty and interference inherent in quantum art, will provide participants with insight into how media affects their behavior [5].

The different ways that music and the arts in general have been extended into the quantum realm have led to the discovery of certain talents that are real qualities. These include contextuality, complementarity, entanglement, and parallelization via coherent superposition, also known as simultaneous linear combination. Let's be clear that the simultaneous existence of several, traditionally mutually incompatible "colors," or visual impressions in general, is what quantum visual art—and specifically quantum parallelism—is all about, not additive color mixing [6].

The fundamental components of matter as well as the superposed immaterial or pre-material states of possibility (or "superposition") in which those components exist before material form are described by quantum physics, which is the foundation of mixed-media artworks [7].

The idea of quantum superposition is initially introduced in an essay, which also highlights how it differs from traditional views of reality. Crooks is able to create a powerful visual representation of the reality of a state of superposition that is invisible to the human eye thanks to the characteristics of digital moving images and the absence of bulky solid elements. The ever-changing yet completely interconnected graphic evokes interconnected, fluid states with constant mobility opportunities [8].

With the use of current quantum computing tools, it is not difficult to experiment with different ideas in the light of creating quantum concept-related arts. With these settings, a new multidisciplinary idea called "quantum art" combines the ideas of quantum mechanics, computational creativity, and visual aesthetics to produce works of art that either visualize quantum phenomena, such as probability amplitudes, entanglement, and superposition, or incorporate quantum-inspired algorithms or quantum computation into the creative process.

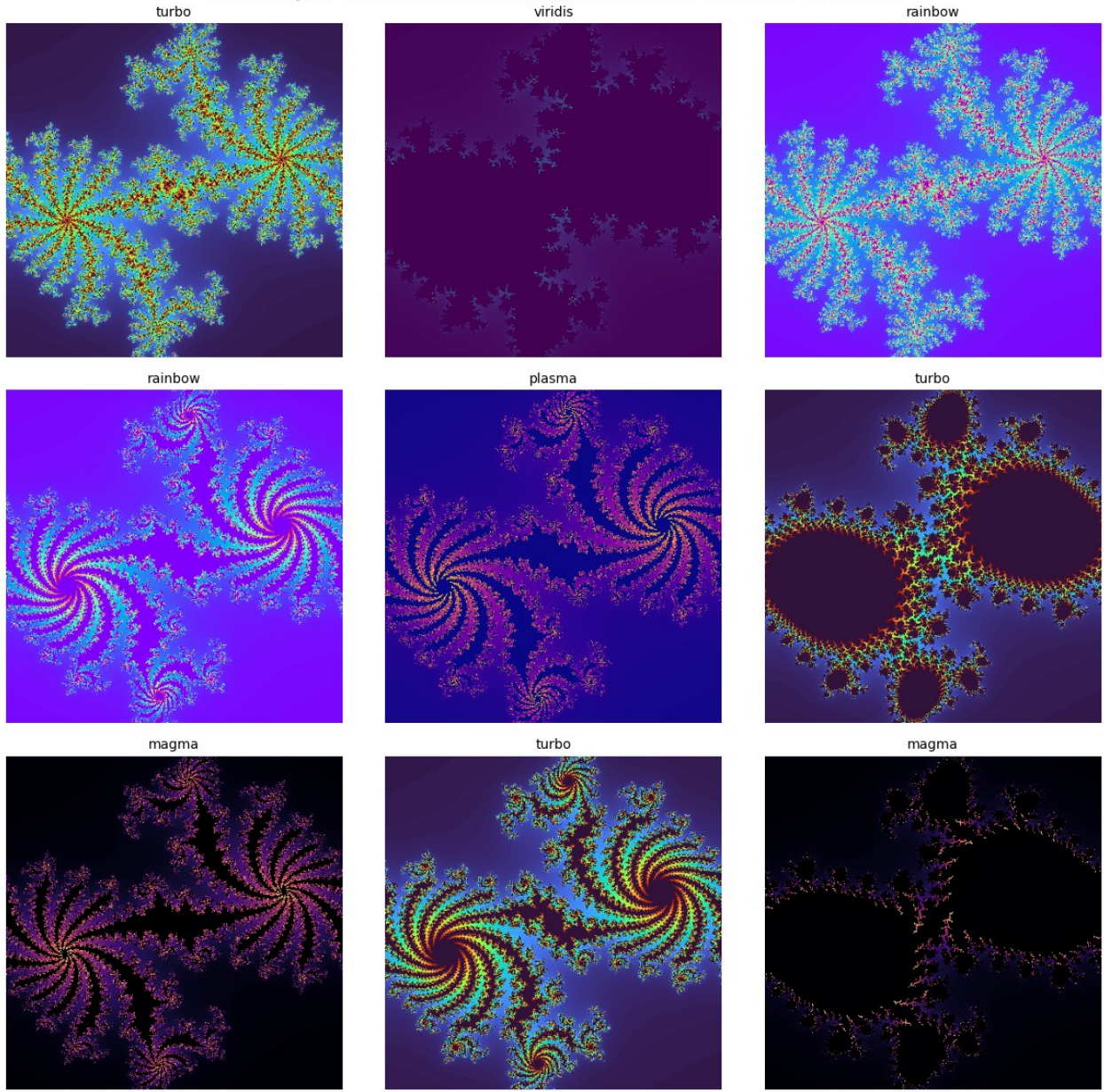

Fig. 1. A sample quantum Julia randomness-derived imageset with a different colormap

Fig. 1 reveals a grid of nine subplots through the Qiskit [9] quantum circuit-based parametrization technique, each of which displays a Julia set fractal created with smooth gradient color maps and quantum randomness. As the Quantum Julia Set Art Series is mentioned in the title, it suggests that these are not entirely deterministic but rather contain some unpredictability derived from probabilistic or quantum-inspired methods. Every subplot has its own visual appeal due to the usage of a different Matplotlib colormap (such as turbo, viridis, rainbow, magma, and plasma).

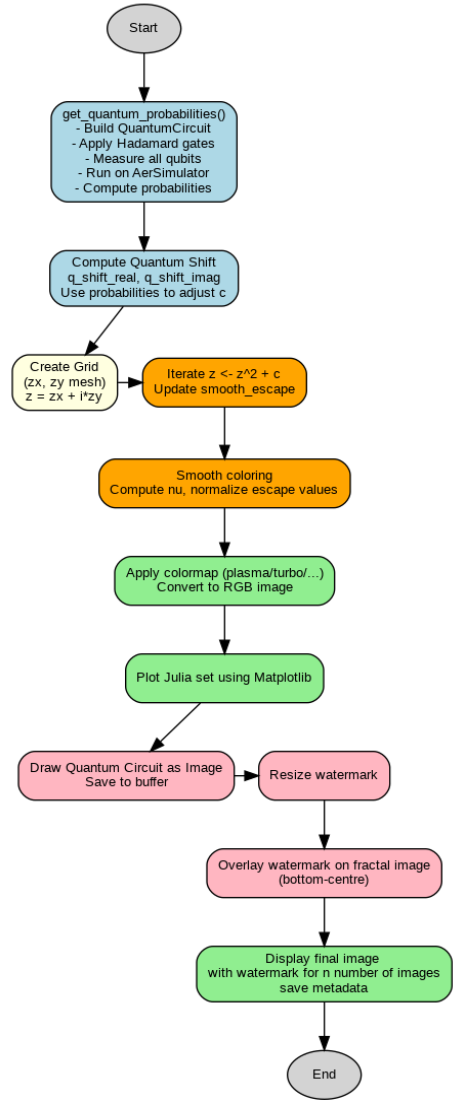

Fig. 2. Flow chart of a quantum circuit-based Julia random image generation algorithm

## II. Background

Researchers and artists are already using quantum computation as an analytical tool or a creative tool [10]. Similarly, AI and quantum experiments were used in data-driven installations such as Refik Anadol's Quantum Memories (NGV Triennial 2020). Refik Anadol created a massive immersive video sculpture by using a neural network on more than 200 million nature photos with Google's quantum computing research (Quantum Supremacy data) [11]. Quantum Memories 2020 is Refik Anadol's most technically and theoretically ambitious piece to date, and the NGV commissioned it. To visualize a dynamic, expansive, immersive multimedia artwork, the piece investigates the possibilities offered by artificial intelligence (AI), machine learning, and quantum computing. To speculate on a different dimension of the natural world as a complex cultural entity with memory, Anadol's work uses a dataset that was gathered from more than two hundred million images related to nature from publicly accessible online resources and processed using quantum computing and machine learning algorithms.

Huffman and Arman combined AI and superposition in their first piece, and entanglement and AI in their second [12]. Yao et al. (2017) showed a quantum algorithm for edge detection and created a framework to encode entire images into qubit states (pixels→amplitudes, positions→basis states) significantly quicker than traditional techniques [13]. Despite being primarily theoretical, these developments demonstrate how quantum algorithms may one day speed up or change computer vision tasks. In a hybrid example, the IQ-PARC group at Northwestern University created a quantum generative model using Microsoft Q# to learn and generate datasets, such as a model of the solar system, that was displayed in three dimensions using Blender and Unreal Engine [14].

In these works, the quantum computation directly drives visual content – often as a probabilistic "generator" of patterns that a human artist then finalizes. Many artists draw directly on quantum ideas (superposition, entanglement, randomness) as conceptual inspiration or formal analogies. The quantum state's indeterminacy and "many possibilities" are treated as metaphors or even literal mechanisms in art-making. In art-making, the indeterminacy and "many possibilities" of the quantum state are regarded as metaphors or even actual mechanisms. As an illustration, Huffman (2021) wrote in an IBM Qiskit blog [15] that the fundamental characteristics of QC (superposition, entanglement, and interference) provide "a new set of affordances" for art, making quantum media as different from traditional digital art as photography is from painting. Quantum physicist-turned-artist Libby Heaney uses live qubits to create images, she explains how to use IBM's quantum computer to generate arbitrary numbers that motivate her "hybrid life-forms" animations (such as her Ent-series), embracing the innate measurement collapse's unpredictable nature [16].

*A. Image Explorer for the Quantum Multiverse*

To create parallel visual representations of the same scene using several quantum measurement bases, drawing inspiration from the interpretation of many worlds. This is aimed at utilizing several unitaries (such as Hadamard, RX, and U3), creating picture states, simulating, and then decoding bitstrings back into image permutations. The image's "universe" varies depending on the base. An apparent Heaney [16] Link envisaged as this expands her concept of parallel worlds into actual visual multiverses, using overlapping biological and technological elements to create "Ent-" beings. This concept could be extended further with the use of quantum algorithms. For example, using Qiskit, a quantum circuit, of the Penrose triangles [17], a generic solution is being aimed to develop using a variety of relevant images with the use of quantum algorithms and circuits for the creation of Impossible objects [18].

As quantum computing unfolds in different domains, in Arts and Humanities, it will transpire inevitably as events are currently being convened [19]. Quantum humanity, a term coined by Barzen [20], design an engaging quantum art-themed 3D animation [21], visual arts and creativity [1], visual arts, and creation, like 3D animation [5], quantum beings [22]. There are some of recent directions coming up in academic settings, quantum-based music software interfacing. How a multidisciplinary, decolonial approach makes use of to explore the fascinating convergence of quantum mechanics and abstract animated visuals, indigenous design features are used to communicate and convey the complex concepts of quantum mechanics. By doing this, traditional views of reality are questioned and rethought [23].

*B. Fractals and Quantum Computing*

Fractals for Understanding Nature [24], [25] since the introduction of concept has increased research and application interests for half a century. Fractals seem to exist at all cosmic scales, including tiny and galaxies and black holes. They fit very nicely with both quantum physics and relativity theories. Numerous scientists have proposed that the structure of the cosmos was fractal [22].

Julia [26] defined the iteration of a quadratic polynomial in the complex plane

$$Z_{n+1} = Z_n^2 + c \quad\quad (1)$$

where:

$z_n$ is a complex number (initial point in the complex plane), c is a complex constant that defines the system's behavior, and n is the iteration step. The sets that emerged from these dynamics are now called Julia sets. Benoît Mandelbrot discovered the Mandelbrot set in 1980's [24]. It is the collection of all complex numbers c that are related to the Julia set. Mandelbrot sets and Julia sets are closely related. Julia's concepts were brought back to life in the 1980s by Douady and Hubbard, who also gave them contemporary rigor and visual impact [27].

How symmetry-protected topological phases of 2D fractal subsystems could be used as resources for universal measurement-based quantum computation has been demonstrated. This is clearly shown for two cluster models that are known to be located within topological phases protected by fractal symmetry, and computationally, it is shown that universality endures at those stages [28].

A new approach that uses differentiable point splatting in conjunction with a bespoke fractal generator to optimize the parameters of the Iterated Function System. Our method successfully traverses the intricate energy landscapes typical of fractal inversion by combining stochastic and gradient-based optimization techniques, guaranteeing strong performance and the capacity to avoid local minima [29].

In SL(2, C), positive matrices can be interpreted physically in two ways: either as Lorentz boosts or as "fuzzy projections" of a spin 1/2 quantum system. In this study, we focus on this second interpretation, following Pertti Lounesto's hints and interpreting them as conformal mappings of the "heavenly sphere" S2 using the traditional Clifford algebraic methods. In the second interpretation, the "boost velocity" is the fuzziness parameter from the first. It is shown that self-similar fractal patterns on S2 result from simple iterative function systems of such maps [30].

The baseline FIC can be accelerated by a factor of using a proposed quantum-accelerated fractal image compression (QAFIC), where and are the numbers of domain blocks and range blocks, respectively. Theoretical analyses and experimental results demonstrate that the suggested QAFIC can significantly lower the time complexity of FIC while preserving the quality of the retrieved photos [31].

Five top fractal applications are as given below [32]:

i) In Fractal cities, cities are those that have a propensity to expand over time in fractal patterns. The pattern created when a huge fractal city absorbs its surrounding towns and villages resembles a self-similar structure that first appears random but is actually a dynamic network that might end up being more effective than contemporary "pre-planned" cities.

ii) In Fractal medicine, understanding fractals is particularly helpful when diagnosing illnesses, such as cancer. Cancerous cells, which develop abnormally, are easier to identify since healthy human blood vessel cells usually grow in an orderly fractal pattern. It is significantly simpler to distinguish between healthy cells and warning indications when using this type of fractal analysis.

iii) In the Resolution and compression of images, Fractal image coding (FIC) and other applications make working with image resolution and even creating 3D models incredibly data-efficient since fractals enable us to transmit seemingly random patterns with less data.

iv) In the Antennas, Antenna design and operation benefit from fractals' generally self-similar characteristics. High-performance, low-profile antennas can be designed using curves such as the Hilbert curve. Multi-band antennas can also be used when they are coupled with ideas about electromagnetic radiation. Go here to learn more.

v) For artists, the breadth of rules governing the production of fractals, from the simple to the complex, is simply irresistible. The Mandelbrot set, for instance, is renowned for offering various "scenes" according to the color scheme chosen for its presentation.

Quantum computing paradigm entails developing quantum circuits [33], which comprises of various gates viz. Hadamard gates, Control-NOT (CNOT), and Rotational gates. Right from the Deutsch algorithm [34], Grover search [35], Shor algorithm [36], Deutsche-Jozsa for rapid solution [37], complexity problem by Berstein Vazirani algorithm [38], the development of quantum circuits and gates for executing quantum computation is fundamental to addressing complicated problems through various principles of quantum physics and mathematical approaches [39].

The aim and objective of this paper are to generate significant fractal images based on a quantum computing approach by using circuits and to analyse the characteristics of the generated dataset with an applicable technique. Quantum circuit-based generation of 1000 fractal images and their metadata, followed by fractal image data analytics, is the contribution of this paper. Using the quantum circuit-based randomness for generating a synthetic fractal image dataset is the novelty of this work.

### III. GENERATION OF QUANTUM FRACTAL VISUAL ARTS

The flowchart, Fig. 2 depicts the steps of generating a quantum circuit-based Julia visual arts image using quantum randomness. Qiskit simulator shots of 2048 for the measurement of probability using the Julia smoothness as a function of width, height and complex constant c, using Eq. 1, have been iterated for the generation of 1000 images with different colormap. The metadata is saved separately in csv file with columns. Post-generation of the images, data analytics including unsupervised technique using K-Means algorithm, was employed after feature extraction for clustering the data in different groups.

Fig. 3 depicts the quantum circuit the picture displays a randomized quantum circuit for four qubits ($q_0$, $q_1$, $q_2$, $q_3$) employed for randomness-based fractal synthesis, quantum feature encoding. Significantly, it entails two distinct characteristics.

1. Quantum Randomness for Julia Parameters

   This step ensures that every fractal created is distinct and impacted by quantum randomness, transforming the image from a deterministic Julia set into a work of art inspired by quantum mechanics. Here, we use quantum probabilities to gently disturb the Julia fractal's shape, which is determined by the constant c. This idea can be used to create quantum-driven chaotic systems or quantum generative art.

2. Fractal Computation (Julia Set)

   It uses the escape-time algorithm, a well-known method for creating fractals. use smoothing (via nu and normalization) instead of blocky iterations to create a high-quality, gradient-colored fractal. Fractal art galleries frequently employ this to provide visual smoothness.

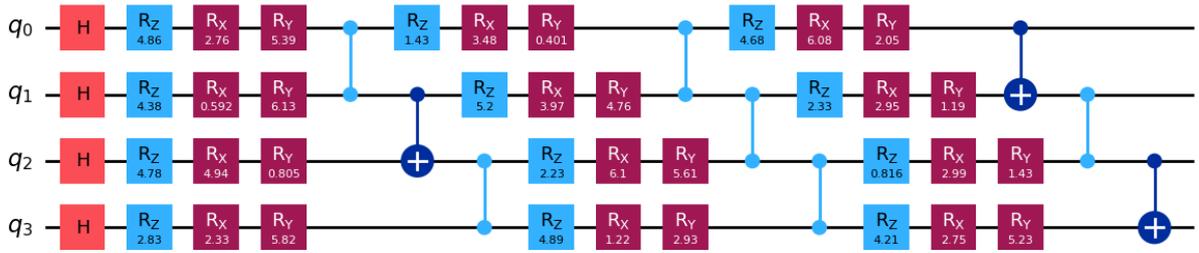

Fig. 3. Four-qubit system quantum circuit with different H, Rx, Ry, Rz, and CNOT gates

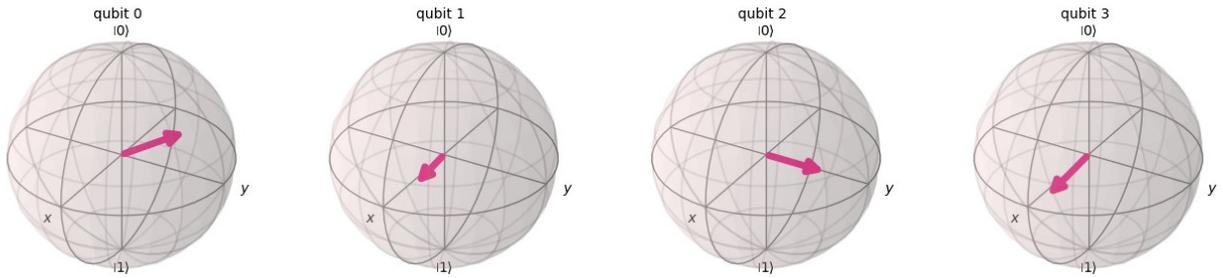

Fig. 4. Four-qubit quantum sample circuit based multti-vector plot

Fig. 4 depicts In the Bloch sphere depiction, one qubit's quantum state is represented by each sphere. The qubit's state vector is shown by the pink arrow. The Axes shown at the top, or north pole, is $|0\rangle$ and $|1\rangle$ at the south pole, or bottom. Superpositions with distinct phases are represented by the x and y axes.

The matrix of the quantum circuit statevector is as given below:

$(0.3547809325+0.0500070229i)|0000\rangle+(-0.1208834684-0.0190024642i)|0001\rangle+(0.2535037569+0.1388187369i)|0010\rangle+(-0.106382878-0.0516752434i)|0011\rangle+(-0.0027190284+0.0806640411i)|0100\rangle+(-0.1091066977-0.2470250053i)|0101\rangle+(-0.271148729-0.0005204541i)|0110\rangle+(0.1104263183-0.1869309506i)|0111\rangle+(0.2297111326+0.2014437232i)|1000\rangle+(-0.1701422265-0.1647323006i)|1001\rangle+(0.147594077+0.116761806i)|1010\rangle+(0.0177752387+0.2213984457i)|1011\rangle+(-0.1074313016+0.1477784289i)|1100\rangle+(-0.0375596472-0.3306229653i)|1101\rangle+(-0.0705529773+0.2592401887i)|1110\rangle+(-0.0875122011-0.317660862i)|1111\rangle$

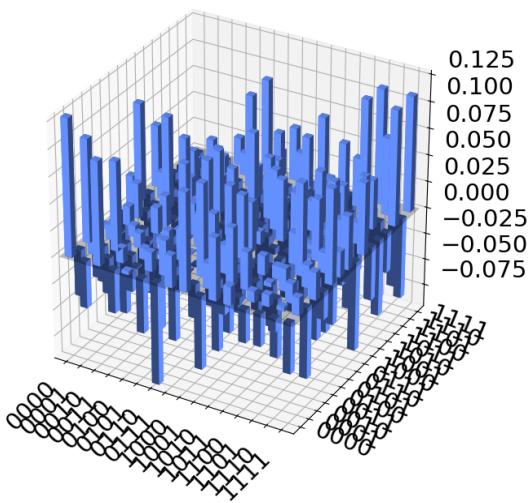
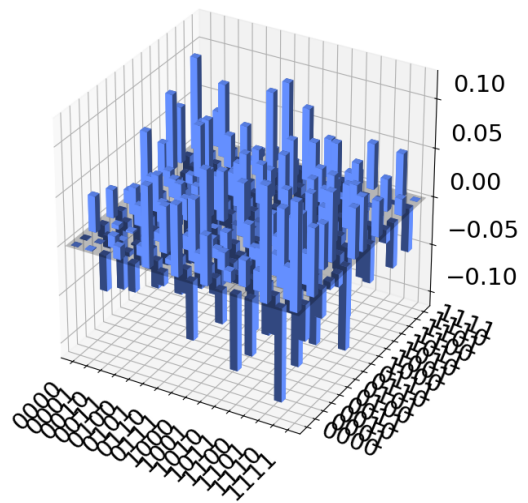

Fig. 5. State city map of a sample quantum circuit of four qubits

Fig. 5 exhibits the real and imaginary parts of the density matrix ($\rho\rho$) of the quantum state generated by the 4-qubit circuit are displayed in this illustration. Strong superposition and entanglement are caused by non-zero off-diagonal elements in both real and imaginary plots. Since the state space is $2^4 = 16$-dimensional, amplitudes disperse over numerous basis states, resulting in modest values (~0.1 max). The many $Rz$, $Rx$, and $Ry$ are suggesting a complicated phase structure, which is indicated by the considerable imaginary portion.

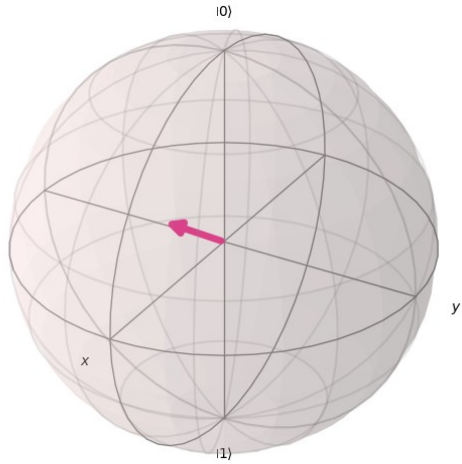

Fig. 6. Bloch vector of the quantum circuit

Plot_bloch_vector() in Qiskit is used to visualize this Bloch sphere, which is the final state of a single qubit following the application of the quantum circuit. The XY-plane position resulted from the last entanglement and rotation operations, which turned it away from the Z-axis and produced a phase shift. This state is a superposition with a relative phase rather than a computational basis state.

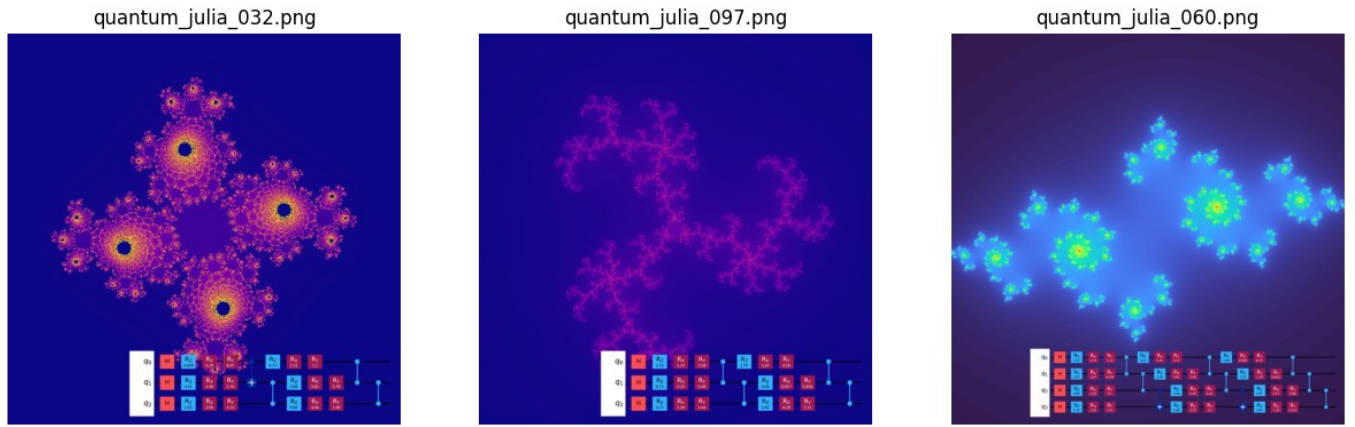

Fig. 7. Sample three generated images from the dataset with different qubit systems and fractal rendering

Fig. 7 reveals three samples from the 1000 samples generated; distinct Julia sets (fractal images) are seen in this image, which was created to investigate quantum-inspired fractal synthesis utilizing parameters based on quantum circuits. Quantum-artistic hybrid representation translates predictable fractal systems to quantum randomness. Data science perspective entails that each image represents a distinct complex function behavior produced from quantum-generated parameters, combining complex dynamics and quantum computing. The discovery of a discrete type implies that the parameter space contains latent structure. Fractal morphology is affected by quantum parameterization.

A. *Fractal Analytics of the generated dataset*

The dataset generated is further put through a data analytics process, which involves missing image data, file name checking, the metadata in CSV format for correctness and visual display, as given below Figs 8 to 16 in conjunction with TABLE I and II.

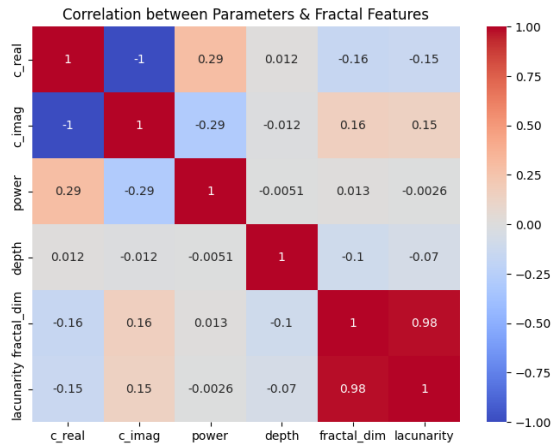

Fig. 8. Correlation heatmap matrix of parameters and fractal features of the generated dataset

Fig. 8 is a correlation heatmap that visualizes the relationships between different parameters and fractal features. c_real, c_imag, power, and depth are the parameters. Features of fractals include lacunarity and fractal_dim (fractal dimension). Encoding colors are Red (close to +1) indicates a strong positive correlation, meaning that as one variable rises, the other one rises as well. Blue (close to -1) as High negative correlation. As one rises, the other falls. Gray (close to 0) implies there is little to no association. C_real and C_imag indicates Correlation = -1 => Perfect negative correlation (perhaps as a result of standardization or dataset design). Given that both are fractal-based measures, the correlation between fractal_dim and lacunarity is 0.98, indicating a very significant positive link. There is a moderately positive association between power and c_real (0.29). There is a moderately negative association (-0.29) between power and c_imag. The association between depth and fractal_dim/lacunarity is very poor, almost zero. These fractal measures are not significantly impacted by iteration depth. The majority of other associations are weak, indicating that fractal metrics are independent of most circuit characteristics (apart from one another).

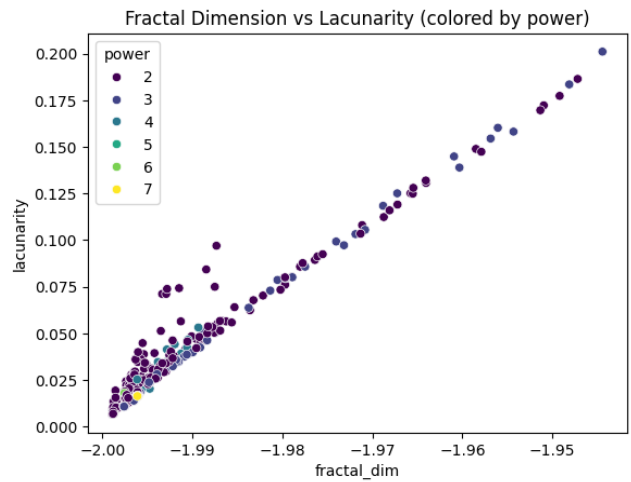

Fig. 9. Fractal dimension and lacunarity plot fo the generated dataset

Fig. 9 reveals with points colored by power, a parameter in the fractal or quantum circuit formation process, this scatter figure illustrates the relationship between fractal dimension and lacunarity. Points are closely spaced on a diagonal, resulting in an almost perfect correlation between lacunarity and fractal dimension which corresponds to the heatmap's 0.98 correlation. Colors show that power has little effect on the linear trend. In contrast to the real/imaginary components of c, higher powers (yellow/green) only slightly alter fractal metrics and do not form a distinct band inside the dense cluster. Fig. 10, 11 and 12 depict the 3D scatter plot, pair-plot and distribution plot of fractal dimension and lacunarity.

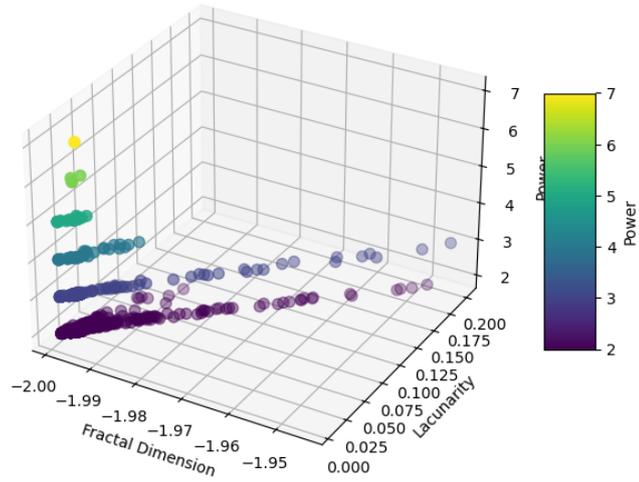

Fig. 10. 3D scatter plot of fractal dimension, lacunarity, and power

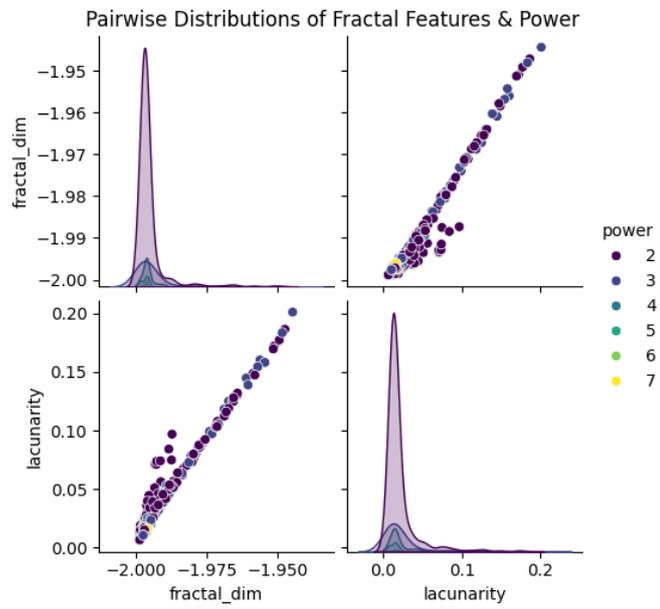

Fig. 11. Fractal features and power pairplot of the generated dataset

The

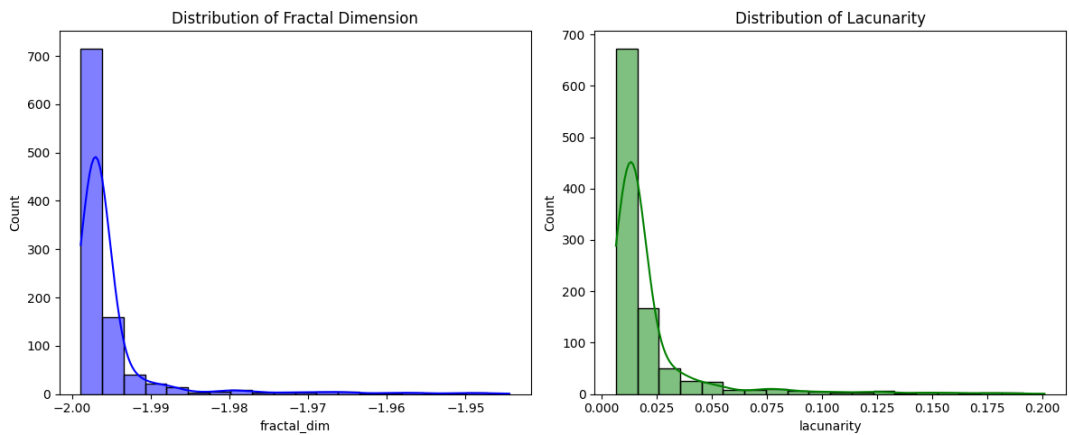

Fig. 12. Distribution plot of fractal dimension and lacunarity

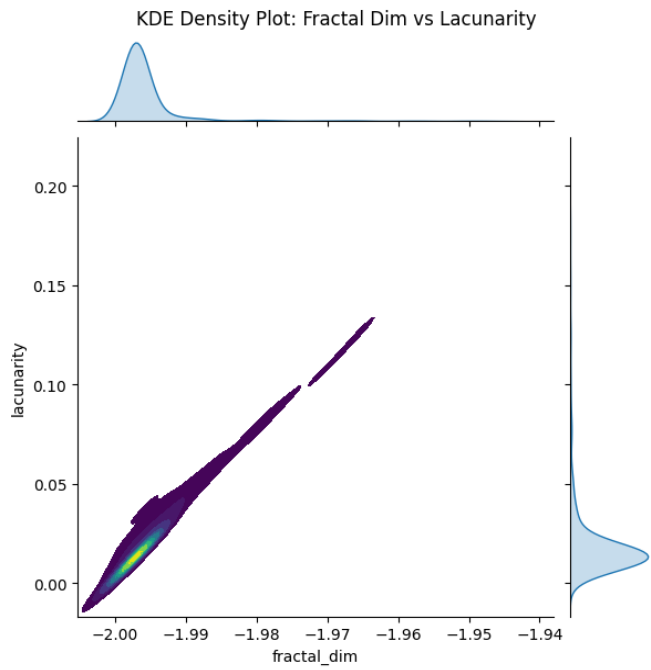

Fig. 13. KDE plot of fractal dimension and lacunarity

The Fig. 13 exhibits the KDE density graphic is visually appealing and demonstrates the close connection between lacunarity and fractal dimension.

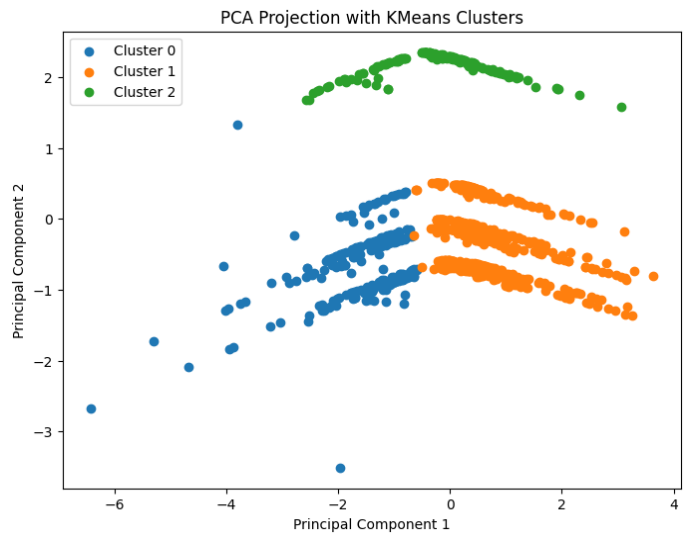

Fig. 14. K-means clustering based on PCA of the multivariate dataset

The extracted features from the generated dataset are further applied through principal component analysis (PCA) for the first two PC. It was then put through the K-Means clustering technique, which yields 3 clusters of the 1000 datasets. Fig. 14 reveals the color-coded clusters on the two PC, whereas Fig. 15 depicts the 3D scatter plot of the clusters in fractal dimension, lacunarity and the power.

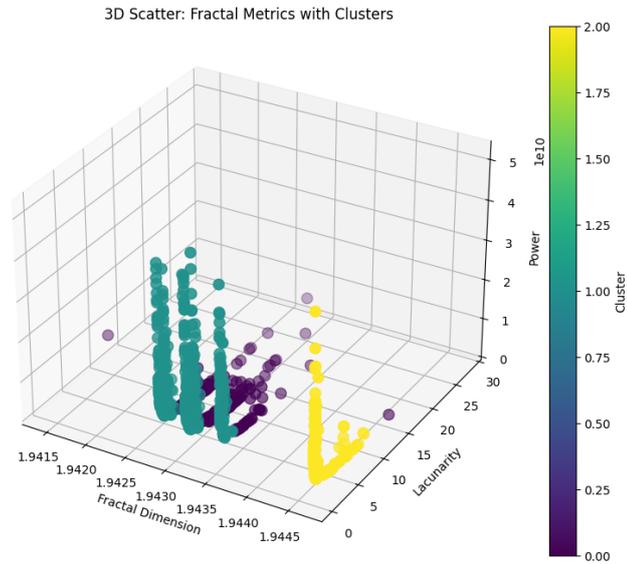

Fig. 15. 3D clustering scatter plot of the three clusters

TABLE I. SAMPLE 3 ROWS OF THE UPDATED METADATA DATAFRAME WITH EXTRACTED FEATURES AND CLUSTER LABELS

| | 612 | 468 | 355 |
|---|---|---|---|
| filename | quantum_julia_613.png | quantum_julia_469.png | quantum_julia_356.png |
| fractal_dim | 1.944661 | 1.943113 | 1.944662 |
| lacunarity | 0.268123 | 0.812675 | 0.472244 |
| power_x | 1.87E+10 | 1.24E+10 | 2.07E+10 |
| seed | 88175571 | 1.9E+09 | 1.6E+09 |
| num_qubits | 3 | 4 | 3 |
| depth | 4 | 2 | 4 |
| c_real | -0.67402 | -0.717432 | -0.72854 |
| c_imag | 0.244173 | 0.287582 | 0.29869 |
| power_y | 2 | 2 | 2 |
| cmap | turbo | viridis | turbo |
| probs_sha1 | e73e173c569eb1e5f179c407f05f1197e035e39c | 325c4813384f7e0e0afcb4f988f069b237f0e268 | 369a11ff08f65ff1485a74adf042fe7ff31afad2 |
| Cluster | 2 | 1 | 2 |
| PC1 | 0.761524 | 0.243373 | 0.887681 |
| PC2 | 2.111231 | -0.227275 | 2.057069 |
| pca1 | 0.761524 | 0.243373 | 0.887681 |
| pca2 | 2.111231 | -0.227275 | 2.057069 |
| cluster_label | 2 | 1 | 0 |

Table I depicts the updated final metadata dataframe including feature extraction and clustering labels of the sample 3 rows of the 1000 images generated during the work course.

TABLE II. CLUSTER PROPERTIES OF THE DATASET

| | | | |
|---|---|---|---|
| **CLUSTER 0:** | Fractal dim=1.943 | Lacunarity=6.802 | Power=6441353940.957 |
| **CLUSTER 1:** | Fractal Dim=1.944 | Lacunarity=0.647 | Power=16330422653.606 |
| **CLUSTER 2:** | Fractal Dim=1.945 | Lacunarity=1.689 | Power=10687890795.811 |

Table II reveals as Cluster 0 (FD≈1.943, Low Power, Lacunarity 6.8). Visual effect as a High gap variation with sparse, tree-like, unconnected branches. Cluster 1 (High Power, Lacunarity 0.647, FD≈1.944). Visual impact is Bright, symmetrical, consistent, dense, and strongly filled. Cluster 2 (FD≈1.945, Medium Power, Lacunarity 1.689). The visual effect is horizontally extended, with some gaps and an intermediate density. Self-similarity: Almost the same FD is shared by all clusters. Heterogeneity detection, as compared to FD, lacunarity distinguishes clusters considerably more strongly. Grouping based on energy/intensity, where Density and power are correlated.

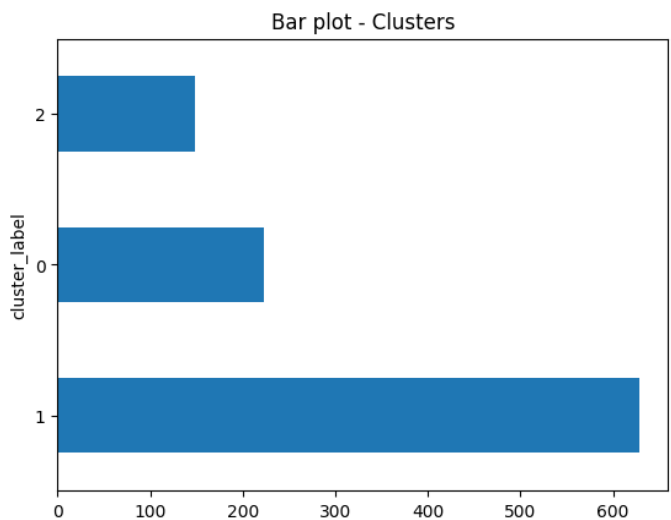

Fig. 16. Bar chart of cluster-based generated dataset

Dominance of Cluster 1 is depicted by the majority of created Julia sets have a homogeneous, dense structure (high power, low lacunarity), which is consistent with Cluster 1's traits. This implies that the fractals produced by the generator are more consistent and dense. Cluster 0 is a minority where the rarity of sparse, dendritic fractals (high lacunarity) may be due to the rarity of such extreme values in the quantum parameter range. The rarest is Cluster 2 when The fact that intermediate density fractals are the least common suggests that the parameter space for these kinds of structures is small. Dense, low-lacunarity structures are favored in the mapping of the parameter space from quantum circuits to fractal parameters. The algorithm could be increasing the frequency of sampling specific areas of the complex plane by the use of a quantum circuit design that does a non-uniform investigation of all parameter areas.

IV. DISCUSSION

Through classical technique, while fractal data generation has been used and finds application in different domains, quantum circuit-based generation offers unique criteria based on quantum-mechanical properties of Nature that reveal quantum randomness in the generated images. Synthetic generation of quantum fractal arts provides an interesting direction to the possible future study of nature's many unfolded mysteries, viz., cosmological conundrums, should it be further studied and applied. Using quantum-circuit-based technique of quantum randomness, 1000 images and their metadata have been further analysed with data analytics and a classical unsupervised classification technique which rendered the dataset as labelled with three classes.

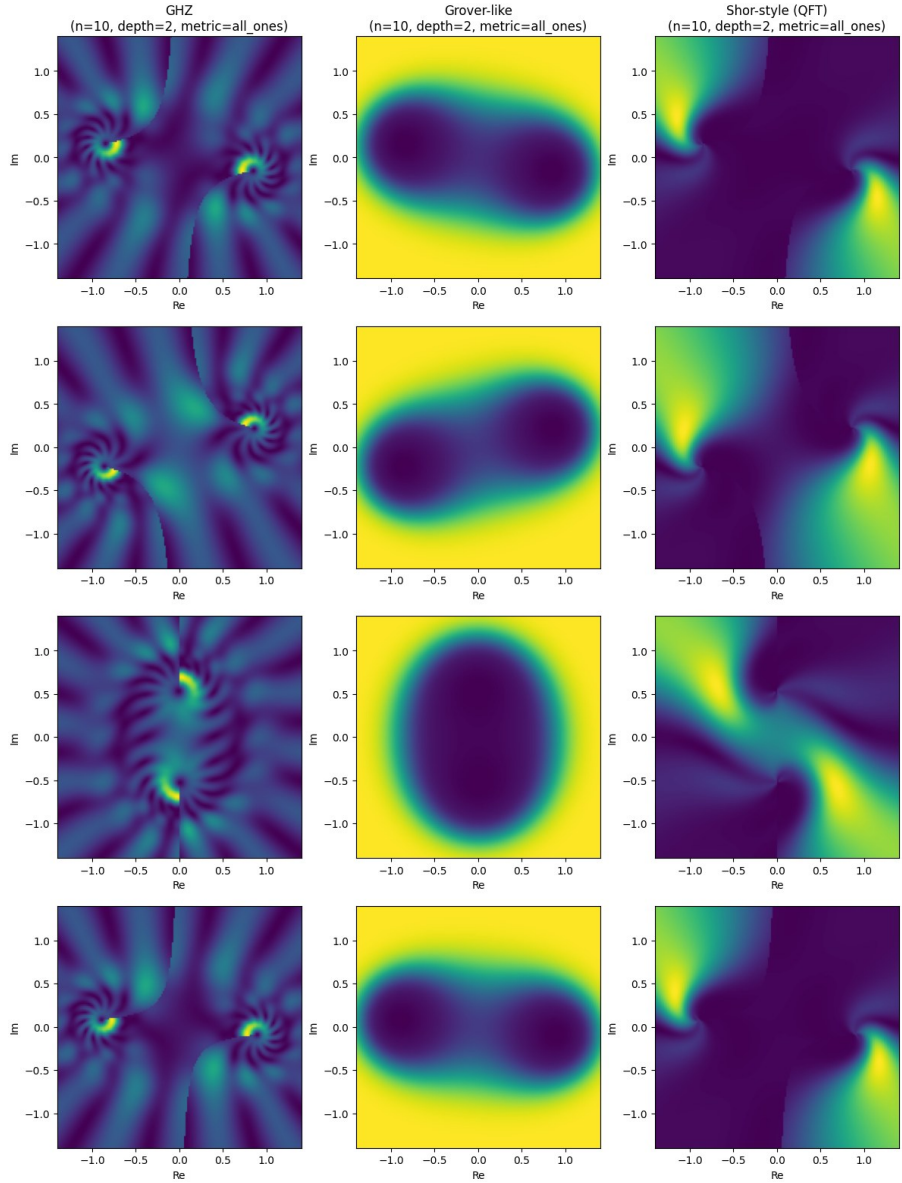

Fig. 17. Using the concept GHZ-Groverlike-Shorlike circuits, Julia generated with 10 qubits

Fig. 17 has been an extension of the concept discussed so far using GHZ, Grover-like, and Shor-like quantum circuits for application in Julia images for different parameters. The parameter landscapes of quantum circuits in the complex plane exhibit organized basins resembling Julia sets, self-similarity, and sensitivity. QFT circuits exhibit periodic structures (rational maps), Grover-like circuits display stable basins (Fatou domains), and GHZ circuits resemble chaotic fractal-like patterns.

## V. Conclusion and Future Direction

The paper describes a quantum circuit-based Julia set dataset generation, which has the underpinning of superposition, randomness, and entanglement. Further, the significant samples generated also provide unique characteristics that the data analytics reveal, classified as three different labels. This upholds the principles of structure discovery in the dataset, homogeneity, algorithmic bias, and objective function optimization, fidelity, dimensional consistency, and NO a priori label. This highlights the potential of using a quantum circuit-based image data generation of different genres for use in different application areas, with manipulation of various parameters.

As an added value of the concept extension, Fig. 17 further resembles the intricacy of Algorithmic Fingerprints, Quantum Complexity & Training Difficulty, Circuit Family Sensitivity, and Landscape Analysis in Complex Parameter Space. This prima facie also appears to be a component of the display of complex-variable quantum landscapes, which are frequently associated with complex variational analysis, fidelity susceptibility, or complex parameter continuation. Analytic continuations of real cost functions into the complex plane are how these quantum landscapes act. Because of their oscillating structure, quantum amplitudes under circuit evolution exhibit patterns that are similar to intricate analytic maps, with interference adding phase sensitivity akin to a fractal.

This technique can be used further for quantum generative arts for different ecosystems with a tailored approach, viz. create a captivating 3D animation with a quantum art concept [21], [40]. Moreover, to complement the difficulty of automatically producing aesthetically pleasing sonifications of paintings by developing a model that makes use of both low-level and high-level visual cues [41]. The work demonstrates that using the Julia-Mandelbrot-Douady-Habbard intertwined concept for quantum circuit-based randomness has the potential to explore novel research directions further.

ACKNOWLEDGMENT

I acknowledge the support and encouragement of everyone who enabled me to develop this article.

REFERENCE

[1] G. Robillard and A. Lioret, "Entangled? Frieder Nake's Probabilities Versus Quantum Computing Artistic Research".

[2] R. P. Feynman, "Simulating Physics with Computers," vol. 21, pp. 467–488, 1982.

[3] E. Schrödinger, *Physique quantique et représentation du monde*. Éditions du Seuil, 1992.

[4] A. Lioret, "Quantum art," *Expressive 2016; Proc. Jt. Symp. Comput. Aesthet. Sketch Based Interfaces Model. Non-Photorealistic Animat. Render.*, pp. 135–139, 2016, doi: 10.2312/exp.20161072.

[5] Z. JIE, S. Dalal, M. S. Bin Abu Bakar, and W. A. Wan Yahaya, "Artistic Creation of Cultural Sustainability and Interactive Creativity through Three-Dimensional Animation," *J. Creat. Arts*, vol. 1, no. 1, pp. 130–144, 2024, doi: 10.24191/jca.v1i1.1475.

[6] V. Putz and K. Svozil, "Quantum music, quantum arts and their perception," in *Quantum Computing in the Arts and Humanities: An Introduction to Core Concepts, Theory and Applications*, Springer, 2022, pp. 179–191.

[7] L. Stone, "Metaphors for Abstract Concepts: Visual Art and Quantum Mechanics," *Stud. Res.*, vol. 1, no. 2, pp. 14–29, 2014.

[8] L. Stone, "Re-Visioning Reality: Quantum Superposition in Visual Art," vol. 51, no. 2, pp. 111–117, 2018, doi: 10.1162/LEON.

[9] IBM Qiskit, "UGate." Accessed: Apr. 20, 2025. [Online]. Available: https://docs.quantum.ibm.com/api/qiskit/qiskit.circuit.library.UGate

[10] A. Crippa, Y. Chai, O. C. Hamido, P. Itaborai, and K. Jansen, "Quantum computing inspired paintings: reinterpreting classical masterpieces," pp. 1–10, 2024, [Online]. Available: http://arxiv.org/abs/2411.09549

[11] "REFIK ANADOL Quantum Memories." [Online]. Available: https://www.ngv.vic.gov.au/refik-anadol-quantum-memories/#:~:text=PROJECT Commissioned by the NGV%2C,work uses the data to

[12] Russell Huffman, "Art by Quantum Entanglement." [Online]. Available: https://medium.com/@jrussellhuffman/art-by-quantum-entanglement-5c3bbafb2400

[13] X. W. Yao *et al.*, "Quantum image processing and its application to edge detection: Theory and experiment," *Phys. Rev. X*, vol. 7, no. 3, pp. 1–14, 2017, doi: 10.1103/PhysRevX.7.031041.

[14] Dongyang Li, "Quantum Generative Art." [Online]. Available: https://iqparc.northwestern.edu/quantum-generative-art/#:~:text=Utilizing a quantum computer and,Blender and Unreal Engine 5

[15] Russell Huffman, "There's A Burgeoning Quantum Art Scene." [Online]. Available: https://medium.com/qiskit/theres-a-burgeoning-quantum-art-scene-76119cca7144

[16] L. Heaney, "The Interview | Libby Heaney," 2025. [Online]. Available: https://www.rightclicksave.com/article/the-interview-libby-heaney#:~:text=In 2021%2C I used one,augmented

[17] R. Penrose, "On the cohomology of Impossible Figures," *Adv. Math. (N. Y).*, vol. 454, pp. 11–16, 2024, doi: 10.1016/j.aim.2024.109868.

[18] "Impossible object - Wikipedia." Accessed: Jul. 20, 2025. [Online]. Available: https://en.wikipedia.org/wiki/Impossible_object

[19] G. Cattan, K. Duś, S. Kusmia, and T. Stopa, "Art makes quantum intuitive," *Front. Quantum Sci. Technol.*, vol. 3, no. May, pp. 1–4, 2024, doi: 10.3389/frqst.2024.1397130.

[20] E. R. Miranda, *Quantum Computing in the Arts and Humanities: An Introduction to Core Concepts, Theory and Applications*. 2022. doi: 10.1007/978-3-030-95538-0.

[21] Z. Jie; and E. Al, *Interaction & Interference: Meditation on Interactive Animation and Media Spectacle in Digital Being*, vol. 1. doi: 10.1007/978-3-031-84636-6.

[22] P. A. Lioret and J. Bouizem, "Fractal Beings," *XX Gener. Art Conf. GA2017*, 2017, [Online]. Available: http://www.stochastic.fr/

[23] M. Chavez and I. Conradi, "Reimagining reality through decolonial and quantum lenses in abstract animation," *Animat. Pract. Process & Prod.*, vol. 12, no. Decolonizing Animation, pp. 137–160, 2023, doi: https://doi.org/10.1386/ap3_00047_1.

[24] B. B. Mandelbrot, *The fractal geometry of nature /Revised and enlarged edition/*. AA(IBM New York), 1983. [Online]. Available: https://ui.adsabs.harvard.edu/abs/1983whf..book.....M

[25] B. B. Mandelbrot, C. J. G. Evertsz, and M. C. Gutzwiller, *Fractals and chaos: the Mandelbrot set and beyond*, vol. 3. Springer, 2004.

[26] GASTON JULIA, "ULIA Mémoire sur l'itération des fonctions rationnelles," vol. 1, pp. 47–245, 1918.

[27] J. H. Hubbard and A. Douady, "Étude dynamique des polynômes complexes," p. 180, 2007.

[28] T. Devakul and D. J. Williamson, "Universal quantum computation using fractal symmetry-protected cluster phases," *Phys. Rev. A*, vol. 98, no. 2, 2018, doi: 10.1103/PhysRevA.98.022332.

[29] A. Djeacoumar, F. Mujkanovic, H. P. Seidel, and T. Leimkühler, "Learning Image Fractals Using Chaotic Differentiable Point Splatting," *Comput. Graph. Forum*, vol. 44, no. 2, 2025, doi: 10.1111/cgf.70084.

[30] A. Jadczyk, "Quantum fractals. Geometric modeling of quantum jumps with conformal maps," *Adv. Appl. Clifford Algebr.*, vol. 18, no. 3–4, pp. 737–754, 2008, doi: 10.1007/s00006-008-0099-2.

[31] S. Du, Y. Yan, and Y. Ma, "Quantum-accelerated fractal image compression: An interdisciplinary approach," *IEEE Signal Process. Lett.*, vol. 22, no. 4, pp. 499–503, 2015, doi: 10.1109/LSP.2014.2363689.

[32] "Top 5 applications of fractals | Mathematics | University of Waterloo." Accessed: Aug. 24, 2025. [Online]. Available: https://uwaterloo.ca/math/news/top-5-applications-fractals

[33] D. . . Deutsch, "Quantum Computational Networks," *Proc. R. Soc. London . Ser. A , Math. Phys. Publ. by R. Soc. Stable*, vol. 425, no. 1868, pp. 73–90, 1989.

[34] D. Aharonov, "Quantum computation," pp. 1–78, 2024.

[35] L. K. Grover, "A fast quantum mechanical algorithm for database search," *Proc. Annu. ACM Symp. Theory Comput.*, vol. Part F1294, pp. 212–219, 1996, doi: 10.1145/237814.237866.

[36] P. W. Shor, "Polynomial-Time Algorithms for Prime Factorization and Discrete Logarithms on a Quantum Computer," vol. 41, no. 2, pp. 303–332, 1999.

[37] D. Deutsch and R. Jozsa, "Rapid solution of problems by quantum computation," *Proc. R. Soc. London. Ser. A Math. Phys. Sci.*, vol. 439, no. 1907, pp. 553–558, 1992.

[38] U. Bernstein, Ethan;, Vazirani, "Quantum Complexity general," 1993, doi: 10.1145/167088.167097.

[39] M. A. Nielsen and I. H. . Chuang, *Quantum Computation and Quantum Information*, 2010th ed. Cambridge University Press.

[40] O. C. Hamido and P. V. Itabora, *OSC-Qasm: Interfacing Music Software with Quantum Computing*. 2023. doi: 10.1007/978-3-031-29956-8.

[41] T. Fink and A. A. Akdag Salah, "Extending the Visual Arts Experience: Sonifying Paintings with AI BT  - Artificial Intelligence in Music, Sound, Art and Design," C. Johnson, N. Rodríguez-Fernández, and S. M. Rebelo, Eds., Cham: Springer Nature Switzerland, 2023, pp. 100–116.